\documentclass[aps,prb,showpacs,twocolumn,groupedaddress]{revtex4}

\usepackage{bm}

\begin{document}

\title{On the thermodynamics of first-order phase transition smeared
by frozen disorder.}

\author{P. N. Timonin}
\email{timonin@aaanet.ru}
\affiliation{Physics Research Institute at Rostov State University
344090, Rostov - on - Don, Russia}

\date{\today}

\begin{abstract}
The simplified model of first-order transition in a media with frozen long-range transition-temperature disorder is considered. It exhibits the smearing of the transition due to appearance of the intermediate inhomogeneous phase with thermodynamics described by the ground state of the short-range random-field Ising model. Thus the model correctly reproduce the persistence of first-order transition only in dimensions $d > 2$, which is found in more realistic models. It also allows to estimate the behavior of thermodynamic parameters near the boundaries of the inhomogeneous phase.
\end{abstract}

\pacs{ 05.70.Jk, 64.60.Cn, 64.60.Fr}

\maketitle

The smearing of first-order phase transitions by frozen disorder of
random bonds (random transition temperature) type is ubiquitous phenomenon, which can be observed in crystalline solid solutions, porous media, gels and composites. The jumps of thermodynamic parameters proper to such transitions could be diminished or completely wiped out by the disorder. According to heuristic criterion derived in Ref. \onlinecite{1} this takes place for sufficiently large range of disorder correlations, greater than the order parameter correlation length and the length defined by the ratio of interphase surface tension and latent heat. The rigorous result was obtained in Ref. \onlinecite{2} for random bond Potts models (RBPM) with first-order transitions in space dimension $d = 2$. It was established that latent heat vanishes in planar Potts models irrespective of disorder strength while long-range order persists at low temperatures. Further numerical studies of q-state RBPM \cite{3,4,5,6} have shown that in $d = 3$ first-order transitions are wiped out only at sufficiently large disorder.

This situation can be compared with that in random field Ising model (RFIM), where, according to Imry-Ma arguments \cite{7} and rigorous results \cite{2, 8},
transition can also exist at $d > 2$ only. This hints on the possibility of RBPM
first-order transitions and RFIM to belong, in some sense, to the same
universality class. This possibility is strongly corroborated by the equivalence of RBPM at $q >> 1$ and RFIM established in Ref. \onlinecite{9} for $d = 2$.
Qualitative arguments in favor of close relation between first-order transitions
in random bond systems and RFIM were advanced in Ref. \onlinecite{10}. Yet
to date there are no analytical results on the thermodynamics of generic smeared
transitions, revealing such relation. So one may try to obtain some insight in this problem considering simplified models.

Here we study the thermodynamics of simple model with long-range
correlated disorder, which may capture the main features of the phenomenology
\cite{1}. According to Ref. \onlinecite{1} the smearing in random
bond (random transition temperature) system results from the appearance of
inhomogeneous equilibrium state, consisting of clusters of two phases (ordered
and disordered), which is energetically more favorable than the homogeneous
one. We find that in present model there is a definite temperature interval where such intermediate inhomogeneous phase exists and its thermodynamics is
described by the RFIM ground state for the strong first-order transition. We obtain the estimates for thermodynamic parameters near the boundaries of this phase.

Let us consider a $d$-dimensional sample undergoing first-order phase transition and divide it on hypercubes of size $L$ in lattice units. We assume that in each hypercube the transition temperature is $T_+$ with probability $p$ and $T_-$ with probability $1-p$, $T_- < T_+$. If $L$ is much larger than the order parameter correlation lengths in the ordered and disordered phases,
\begin{equation}
L \gg \max \left( {\xi _1 ,\xi _2 } \right) \ge 1,
\label{eq:1}
\end{equation}
thermodynamics of each cube can be described by the density of inequilibrium thermodynamic potential $f\left( {\bm\varphi ,\tau _ \pm  } \right)$
 of infinite sample. Here $\bm\varphi$ is multicomponent order parameter and $
\tau _ \pm   \equiv T/T_ \pm   - 1$  denotes the reduced temperatures, which correspond to the regions with transitions at $T_+$ and $T_-$ . So the effective Hamiltonian of the system can be expressed as
\begin{equation}
{\cal H}\left( {\left\{\bm\varphi  \right\},{\bm\sigma }} \right) = L^d \sum\limits_{i = 1}^N {f\left( {\varphi _i ,\tau _{\sigma _i } } \right)}  + {\cal H}_{{\mathop{\rm int}} } \left( {\left\{\bm\varphi  \right\}} \right).
\label{eq:2}
\end{equation}
with random variables $\sigma _i  =  \pm $, $N$ being a number of cubic blocks and ${\cal H}_{{\mathop{\rm int}} } \left( {\left\{\bm \varphi  \right\}} \right)$ representing the surface interaction of neighboring blocks.  The interaction term is proportional to $L^{d - 1} $ and it should depend on the difference of $\bm\varphi$ in the nearby cubes. So we put
\begin{equation}
{\cal H}_{{\mathop{\rm int}} } \left( {\left\{\bm \varphi  \right\}} \right) = L^{d - 1} A\sum\limits_{\left\langle {ij} \right\rangle } {\bm\left( {\bm\varphi _i  - \bm\varphi _j } \right)^2 },
\label{eq:3}
\end{equation}
where $A$ is some constant.
 
The density of the average equilibrium potential of the model is
\begin{equation}
\left\langle f \right\rangle  =  - N^{ - 1} L^{ - d} T\left\langle {\ln \int {d \bm\varphi \exp \left[ { - \beta {\cal H}\left( {\left\{ \bm\varphi  \right\},{\bm \sigma }} \right)} \right]} } \right\rangle _\sigma  
\label{eq:4}
\end{equation}
As $L \gg 1$ integral in Eq. (\ref{eq:4}) can be estimated via the method of steepest descend. So we must find a global minimum of ${\cal H}\left( {\left\{ \bm\varphi  \right\},{\bm \sigma }} \right)$. Here we assume that at temperatures including the interval $T_ -   < T < T_ +  $ both $f\left( {\bm\varphi ,\tau _ +  } \right)$ and $f\left( {\bm\varphi ,\tau _ -  } \right)$
 have minimum at $\bm\varphi = 0$ and $q$ minima at $\bm\varphi  = \bm\varphi _\alpha  \left( {\tau _ +  } \right)$ or $\bm\varphi  = \bm\varphi _\alpha  \left( {\tau _ -  } \right)$, $\alpha  = 1,2,...,q$, correspondingly. We also suppose, that $f\left( {\bm\varphi ,\tau } \right)$ is invariant under some symmetry group, so for all $\alpha$
\[
\varphi _\alpha  \left( \tau  \right) = \varphi _s \left( \tau  \right),\\
\qquad
f\left( {\bm\varphi _\alpha  ,\tau } \right) = f_s \left( \tau  \right)
\]
and $\sum\limits_{\alpha  = 1}^q {\bm\varphi _\alpha  }  = 0$.
 
 Then ${\cal H}\left( {\left\{ \bm\varphi  \right\},{\bm\sigma }} \right)$
 has $\left( {q + 1} \right)^N $ local minima at which every $\bm\varphi _i $
 is close to one of the $f\left( {\bm\varphi ,\tau _{\sigma _i } } \right)$
 minima differing from them by terms of order $1/L$. 
These local minima can be parameterized by the vectors $\bm\psi _i $, which take values $\bm\varphi _\alpha  /\varphi _s $ and 0, so we have at them
\[
\bm\varphi _i  = \bm\psi _i \varphi _s \left( {\tau _{\sigma _i } } \right) + O\left( {1/L} \right).
\]

For the strong first order transition we may assume that at $T_ -   < T < T_ +  $
\[ 
\varphi _s \left( {\tau _ +  } \right) \approx \varphi _s \left( {\tau _ -  } \right) \equiv \varphi _s.
\]
Then we get for the value of Hamiltonian at local minimum
\begin{eqnarray}
{\cal H}\left( {\left\{ \bm\psi  \right\},{\bm \sigma }} \right) = L^d \sum\limits_{i = 1}^N {\left[ {f_0 \left( {\tau _{\sigma _i } } \right) + \psi _i \delta f\left( {\tau _{\sigma _i } } \right)} \right]} \nonumber \\
 + 2L^{d - 1} A\varphi _s^2 \sum\limits_{\left\langle {ij} \right\rangle } {\left( {\psi _i  - \bm\psi _i \bm\psi _j } \right)} 
\label{eq:4a} 
 \end{eqnarray}

Here $f_0 \left( {\tau _\sigma  } \right) \equiv f\left( {0,\tau _\sigma  } \right)$, $\delta f\left( {\tau _{\sigma } } \right) \equiv f_s \left( {\tau _{\sigma } } \right) - f_0 \left( {\tau _{\sigma } } \right)$
 and we used the relation $\psi _i^2  = \psi _i $ being the consequence of $\psi _i  = \left\{ {0,1} \right\}$. So we have
\[ 
\left\langle f \right\rangle  = N^{ - 1} L^{ - d} \left\langle {\mathop {\min }\limits_{\left\{\bm\psi  \right\}} {\cal H}\left( {\left\{ \bm\psi  \right\},{\bm \sigma }} \right)} \right\rangle _\sigma  
\]

Apparently, the global minimum is realized on the configurations, which have parallel $\bm\psi _i $ and $\bm\psi _j $ on neighboring sites (if they are both nonzero). Limiting the choice to such configurations, we can put $\bm\psi _i \bm\psi _j  = \psi _i \psi _j $  in Eq. (\ref{eq:4a}). Then, expressing $\psi_i$ via Ising spins $s_i  =  \pm 1$, $\psi _i  = \left( {1 - s_i } \right)/2$, we get
\begin{eqnarray}
\left\langle f \right\rangle  = \frac{1}{2}\left[ {\left\langle {f_0 \left( {\tau _\sigma  } \right)} \right\rangle _\sigma   + \left\langle {f_s \left( {\tau _\sigma  } \right)} \right\rangle _\sigma  } \right] + dJ \nonumber \\
+ N^{ - 1} \left\langle {\mathop {\min }\limits_{\bm s} E\left( {{\bm s},{\bm\sigma }} \right)} \right\rangle _\sigma,
\label{eq:5}
\end{eqnarray}
\begin{eqnarray*}
E\left( {{\bm s},{\bm \sigma }} \right) =  - \sum\limits_{i = 1}^N {H_i s_i  - \frac{J}{2}} \sum\limits_{\left\langle {ij} \right\rangle } {s_i s_j },
\\
H_i  = \delta f_{\sigma _i } /2,
\qquad
J = A\varphi _s^2 /L.
\end{eqnarray*}

Thus for large disorder correlation length $L$ and strong first-order transition the average equilibrium potential of this simple random-temperature system is determined by the ground state of short-range Ising model with random fields having bimodal distribution.

The present results rely on the large disorder correlation length $L$ and the
 possibility to ignore the order parameter temperature dependence for strong
 first-order transitions. Actually it could be sufficient to require (along
with the condition (\ref{eq:1})) the first-order transition to be strong with
 small order parameter fluctuations around the local minima. Indeed, Eq.
 (\ref{eq:5}) holds, when
\begin{equation}
L_c^{ - d}  \equiv \beta \varphi _s^2 \min \left[ {f''_{min}\left( {0,0} \right), f''_{min}\left( {\varphi _s ,0} \right)} \right] \gg L^{ - d} ,
\label{eq:5a}
\end{equation}
where $f''_{min}\left( {\varphi ,\tau} \right)$ is the lowest eigenvalue of the  matrix of the second-order $\bm\varphi$-derivatives of the potential. So, if $L_c  \le \max \left( {\xi _1 ,\xi _2 } \right)$ (meaning small order parameter fluctuations), the condition (\ref{eq:5a}) does not impose further restriction on $L$ as compared with that of Eq. (\ref{eq:1}). Also small $L_c $ implies only slight temperature dependence of $\varphi _s \left( {\tau _ \pm  }\right)$. In case of a soft transition, $L_c  \gg \max \left( {\xi _1 ,\xi _2 } \right)$, and for $\max \left( {\xi _1 ,\xi _2 } \right) \ll L \le L_c $ one should take into account the fluctuations around the local minima, as well as $\varphi _s \left( {\tau _ \pm  } \right)$ temperature dependence, which would essentially modify the effective spin Hamiltonian breaking the simple RFIM equivalence.

To find the RFIM ground state in Eq. (\ref{eq:5}) is nontrivial task only in the interval $T_ -   < T < T_ +  $ owing to the definite temperature dependence of two possible $H_i$ values. Indeed, $H_+ =\delta f_ +/2$ and $H_- = \delta f_-/2$ grow monotonously with increasing $T$ and change sign at $T_+$ and $T_-$
correspondingly. Particularly at small $\left| {\tau _ \pm  } \right| <  < 1$
we have
\[
\delta f_ \pm   \approx T_ \pm  \tau _ \pm  \delta S,
\]
where $\delta S > 0$ is the entropy jump at the transition in either component.
So at $T > T_ + $ all $H_i  > 0$  and ground state is $s_i  =  + 1$ for all $i$ and
\[
\left\langle f \right\rangle  = \left\langle {f_0 (\tau _\sigma  )} \right\rangle
_\sigma   = pf_0 (\tau _ +  ) + \left( {1 - p} \right)f_0 (\tau _ -  ),
\]
while  at $T < T_ - $  all  $H_i  < 0$ , so ground state is $s_i  =  - 1$  and
\[
\left\langle f \right\rangle  = \left\langle {f_s (\tau _\sigma  )} \right\rangle_\sigma.
\]

Thus above $T_+$ all cubic blocks are in the disordered phase and equilibrium
potential is that of this phase averaged over random transition temperature.
Similarly below $T_-$ we have all blocks in the ordered state. So in the interval $T_ -   < T < T_ +  $ gradual transformation from homogeneous disordered state to a homogeneous ordered one takes place. Inside this interval the remnant jumps of first-order transition can exist at some $T_0$ defined by the condition $\left\langle {H_i } \right\rangle _\sigma   = 0$ or
\[
\left\langle {f_0 (\tau _\sigma  )} \right\rangle _\sigma   = 
\left\langle {f_s (\tau_\sigma  )} \right\rangle _\sigma .
\]

The present results can be trivially generalized to the case when random 
transition temperatures are continuously distributed between $T_-$ and $T_+$, 
which would give continuous $H_i $ distribution with bounded support. Yet, 
according to numerical studies \cite{11,12,13,14}, thermodynamics of RFIM at 
$T=0$ does not crucially depend on the random field distribution, apart from 
the macroscopic ground state degeneracy in the bimodal case \cite{15}. 
As generally the long-range order in the RFIM ground state exists only in 
$d \ge 3$ (for small disorder), we have the same condition for the persistence 
of first-order transition in the present model for generic transition 
temperature disorder. So it is similar in this respect to the more realistic 
RBPM \cite{2,3,4,5,6,9,10,16}.

Yet there is apparent qualitative discrepancy with numerical results for RBPM, 
which indicate that in $d = 2$ first-order transition transforms into a 
second-order one \cite{9,10,16}, while the study of 2d RFIM ground state in 
homogeneous field \cite{13} shows that there is no transition at 
$\left\langle {H_i } \right\rangle _\sigma   = 0$  ($T = T_0$). Thus present 
model cannot elucidate the origin of the instability and the nature of the 
order parameter appearing at $T_0$ in $2d$ RBPM. Still we may suppose that far 
from $T_0$ RFIM ground state correctly reproduce the qualitative features of the inhomogeneous equilibrium state in realistic models, especially the existence 
of intermediate inhomogeneous phase in definite temperature interval 
$T_ -  < T < T_ +  $ in which random transition temperatures vary. 
Actually, the Imry-Wortis phenomenology \cite{1} already implies its existence, 
but to date numerical \cite {3,4,5,6,16} and renormalization group \cite{9,10,17} studies of RBPM were not intended to reveal such phase.  

We can give rough estimates for the behavior of thermodynamic parameters
near $T_-$ and $T_+$ inside the inhomogeneous phase. Thus near $T_+$
negative fields $H_ + = \delta f_ +  /2$  are small so in the sea of positive spins only very large (and, hence, very rare) clusters of negative spins (ordered phase) can have sufficiently large energy gain as compared to the surface energy loss to diminish RFIM ground state energy. Then we can choose approximate trial ground state dividing the lattice on the cubic blocks of size $\Lambda  >  > 1$ and putting all spins positive except for the cubes in which homogeneous component of random field $H_c  \equiv \Lambda ^{ - d} \sum\limits_{i \in c} {H_i }  <  - dJ/\Lambda $. In the last blocks we put all $s_i  =  - 1$. As there is small probability to find nearby two or more such blocks, we get for the energy of such spin configurations in the most of the disorder realizations
\begin{eqnarray}
E\left( {\Lambda ,{\bf \sigma }} \right) =  - NdJ - \sum\limits_{i = 1}^N {H_i }
\nonumber
\\
+ \Lambda ^d \sum\limits_{c = 1}^{N/\Lambda ^d } {\left( {2H_c  +
dJ/\Lambda } \right)\vartheta \left( { - H_c  - dJ/\Lambda } \right)}.
\label{eq:6}
\end{eqnarray}
As
\[
%\text{As }
\left\langle {\mathop {\min }\limits_{\bf s} E\left( {{\bf s},{\bf \sigma }}
\right)} \right\rangle _\sigma   \le \left\langle {\mathop {\min }\limits_\Lambda
E\left( {\Lambda ,{\bf \sigma }} \right)} \right\rangle _\sigma   \le \mathop
{\min }\limits_\Lambda  \left\langle {E\left( {\Lambda ,{\bf \sigma }} \right)}
\right\rangle _\sigma  ,
\]
we can get upper bound for the ground state energy calculating $\mathop {\min
}\limits_\Lambda  \left\langle {E\left( {\Lambda ,{\bf \sigma }} \right)}
\right\rangle _\sigma  $.
From Eq. (\ref{eq:6}) it follows
\begin{eqnarray}
\mathop {\min }\limits_\Lambda  \left\langle {E\left( {\Lambda ,{\bf \sigma }}
\right)} \right\rangle _\sigma  /N =  - dJ - \left\langle {H_i } \right\rangle
_\sigma   + U_+. \label{eq:7}
\\
U_ +   \equiv \mathop {\min }\limits_\Lambda  \left[ {\int\limits_{H_ +  }^{ -
dJ/\Lambda } {dHW\left( H \right)\left( {2H + dJ/\Lambda } \right)} } \right]
\label{eq:8}
\end{eqnarray}
Here $W\left( H \right)$ is the distribution function for homogeneous field
$H_c  \equiv \Lambda ^{ - d} \sum\limits_{i \in c} {H_i } $
 in blocks considered. For $H$ close to $H_+$ and $\Lambda  >  > 1$
 it has the form
\[
W\left( H \right) = \left[ {\frac{{\Lambda ^d }}{{2\pi \left( {H_ -   - H_ +  }
\right)\left( {H - H_ +  } \right)}}} \right]^{1/2} p^{\Lambda ^d }.
\]
The minimum of the integral in Eq. (\ref{eq:8}) is attained at the size
\[
\Lambda _ +   \approx \frac{{dJ}}{{\left| {H_ +  } \right|}}\left[ {1 +
\frac{{\left| {H_ +  } \right|^d }}{{2d\ln \left( {1/p} \right)\left( {dJ} \right)^d
}}} \right],
\]
which diverges when $T \to T_ +   - 0$. Thus we have near $T_+$
\begin{eqnarray}
 U_ +   \approx  - \left( {\frac{{\left| {H_ +  } \right|^3 }}{{\pi d\ln \left( {1/p}
\right)H_ -  }}} \right)^{1/2} p^{\Lambda _ + ^d } \approx \nonumber
\\
 - \left( {\frac{{\left| {\delta f_ +  } \right|^3 }}{{4\pi d\ln \left( {1/p}
\right)\delta f_ -  }}} \right)^{1/2} \exp \left[ { - \ln \left( {1/p} \right)\left(
{\frac{{2dJ}}{{\left| {\delta f_ +  } \right|}}} \right)^d } \right]\label{eq:9} \end{eqnarray}

 This expression describes, apparently, the contribution of very large and very
rare clusters of ordered phase appearing immediately below $T_+$, which
diminish the energy. The probability to find such clusters, $p^{\Lambda _ + ^d
} $, vanishes very fast at $T_+$ (as $\exp \left( { - const/\left| {\tau _ +  } \right|^d } \right)$) and it dominates the temperature behavior of $U_+$. On physical grounds, the presence of this dominating term in Eq. (\ref{eq:9}) can be expected also in the true ground state energy, so we may assume that expression in Eq. (\ref{eq:7}) is rather close to it. So we suppose that near $T_+$
\[
\left\langle f \right\rangle  \approx \left\langle {f_1 \left( {\tau _\sigma  }
\right)} \right\rangle _\sigma   + U_ +  .
\]
Then average entropy and heat capacity are
\begin{eqnarray*}
\left\langle S \right\rangle  \approx \left\langle {S_0 \left( {\tau _\sigma  }
\right)} \right\rangle _\sigma   - \frac{{\left| {U_ +  } \right|}}{{2J}}\left(
{\frac{{2dJ}}{{\left| {\delta f_ +  } \right|}}} \right)^{d + 1} \delta S\ln \left(
{1/p} \right).
\\
\left\langle C \right\rangle  \approx \left\langle {C_0 \left( {\tau _\sigma  }
\right)} \right\rangle _\sigma   + \frac{{\left| {U_ +  } \right|T_ +  }}{{4J^2
}}\left( {\frac{{2dJ}}{{\left| {\delta f_ +  } \right|}}} \right)^{2d + 2} \delta S^2 \ln ^2 \left( {1/p} \right)
\\
\text{Here } S_0 \left( \tau  \right) \equiv  - \frac{{\partial f_0 \left( \tau
\right)}}{{\partial T}},
\qquad
C_0 \left( \tau  \right) \equiv  - T\frac{{\partial ^2 f_0
\left( \tau  \right)}}{{\partial T^2 }}.
\end{eqnarray*}
Thus slight diminishing of the entropy starts already at $T_+$ indicating the rounding of the transition. Accordingly, the heat capacity becomes larger than its mean-field value below $T_+$.

In order to find the spontaneous order parameter, one should consider a system in an infinitesimal field $\bm h$ conjugate with the order parameter. In the presence of such field we should take into account only one nontrivial minimum of the potential, namely, the one with $\bm\varphi_{\alpha}$ direction closest to that of $\bm h$ as the others have higher values. Let it be the minimum at $\bm\varphi_1$. Then two values of spins in Eq. (\ref{eq:5}) correspond to $\bm\varphi =0$ ($s=+1$) and $\bm\varphi =\bm\varphi_1$ ($s=-1$), so the average spontaneous order parameter is
\[
\left\langle \bm\varphi  \right\rangle  = \frac{\bm\varphi _1}{2} \left[{1 - N^{ - 1} \sum\limits_i {\left\langle s_i^{\left( 0 \right)}\left(\bm\sigma\right)\right\rangle_\sigma } } \right]
\]
Here $ s_i^{\left( 0 \right)}\left(\bm\sigma\right)$ is ground state spin configuration. When there are several ground states the average over all of them should be taken in this expression. Using the above trial ground state with $\Lambda = \Lambda_+$ we get
\begin{equation}
\left\langle \bm\varphi  \right\rangle  = \bm\varphi_1 \int\limits_{H_ +  }^{ -
dJ/\Lambda_+ } {dHW\left( H \right)} \approx \bm\varphi_1 \left|U_+/H_+\right|
\label{eq:10}
\end{equation}
Thus spontaneous order parameter appears continuously below $T_+$. Yet for every disorder configuration the transition into inhomogeneous phase near this point is a first-order one. The absence of discontinuities at $T_+$ in the average thermodynamic parameters results from the vanishing probability to have the finite jumps in all of them including $\left\langle \bm\varphi \right\rangle$. 

In the vicinity of $T_-$ , where rare clusters of disordered phase exist,
analogous treatment gives the following estimates
\[
\left\langle f \right\rangle  \approx \left\langle {f_s \left( {\tau _\sigma  }
\right)} \right\rangle _\sigma   - U_ -  ,
\]
\[
\left\langle \bm\varphi  \right\rangle  \approx \bm\varphi _1\left(1 - U_-/H_- \right)
\]
\begin{eqnarray*}
\left\langle S \right\rangle  \approx \left\langle {S_s \left( {\tau _\sigma  }
\right)} \right\rangle _\sigma   + \frac{{U_ -  }}{{2J}}\left(
{\frac{{2dJ}}{{\left| {\delta f_ +  } \right|}}} \right)^{d + 1} \delta S\ln \left({\frac{1}{{1 - p}}} \right).\\
\left\langle C \right\rangle  \approx \left\langle {C_s \left( {\tau _\sigma  }
\right)} \right\rangle _\sigma   + \frac{{U_ -  T_ -  }}{{4J^2 }}\left(
{\frac{{2dJ}}{{\delta f_ -  }}} \right)^{2d + 2} \delta S^2 \ln ^2 \left(
{\frac{1}{{1 - p}}} \right)\\
U_ -   = \left( {\frac{{\delta f_ - ^3 }}{{4\pi d\ln \left( {1 - p} \right)\delta f_ +}}} \right)^{1/2} \exp \left[ {\ln \left( {1 - p} \right)\left( {\frac{{2dJ}}{{\delta f_ -  }}} \right)^d } \right]
\end{eqnarray*}
Thus near $T_-$ there also are exponentially small contributions (proportional to $\exp \left( { - const/\tau _ - ^d } \right)$) to the mean-field values of
thermodynamic parameters, which indicate the softening of the transition at $T_0$.

We should note that the slight anomalies at the boundaries of
inhomogeneous phase are hard to observe in the numerical simulations. As they
result from the rare appearance of very large clusters of opposite phase, the
large samples and large number of disorder realizations are needed to reveal
these anomalies at $T_+$ and $T_-$. Yet the tails in the temperature
dependencies of thermodynamic parameters, similar to those described by Eq.
(\ref{eq:10}), are often seen in experimental studies of first-order transitions in random media (see, for example, Ref. \onlinecite{18}). This may indicate the
presence of intermediate inhomogeneous phase in real systems.

This may also imply that the present model with seemingly unrealistic disorder is more closely related to the realistic models than one may expect. Indeed, one can imagine that application to, say, RBPM of some sort of renormalization group procedure, which eliminates the order parameter fluctuations on scales smaller than some large $L$, will result in effective Hamiltonian, similar to that in Eq. (\ref{eq:1}) in some range of the model parameters.  
\begin{acknowledgments}
This work was made under support from INTAS, grant 2001-0826.
I gratefully acknowledge useful discussions with V. P. Sakhnenko, V. I. Torgashev and V. B. Shirokov.
\end{acknowledgments}


\begin{thebibliography}{18}
\expandafter\ifx\csname natexlab\endcsname\relax\def\natexlab#1{#1}\fi
\expandafter\ifx\csname bibnamefont\endcsname\relax
  \def\bibnamefont#1{#1}\fi
\expandafter\ifx\csname bibfnamefont\endcsname\relax
  \def\bibfnamefont#1{#1}\fi
\expandafter\ifx\csname citenamefont\endcsname\relax
  \def\citenamefont#1{#1}\fi
\expandafter\ifx\csname url\endcsname\relax
  \def\url#1{\texttt{#1}}\fi
\expandafter\ifx\csname urlprefix\endcsname\relax\def\urlprefix{URL }\fi
\providecommand{\bibinfo}[2]{#2}
\providecommand{\eprint}[2][]{\url{#2}}

\bibitem[{\citenamefont{Imry and Wortis}(1979)}]{1}
\bibinfo{author}{\bibfnamefont{Y.}~\bibnamefont{Imry}} \bibnamefont{and}
  \bibinfo{author}{\bibfnamefont{M.}~\bibnamefont{Wortis}},
  \bibinfo{journal}{Phys.\ Rev.\ B} \textbf{\bibinfo{volume}{19}},
  \bibinfo{pages}{3580} (\bibinfo{year}{1979}).

\bibitem[{\citenamefont{Aizenman and Wehr}(1989)}]{2}
\bibinfo{author}{\bibfnamefont{M.}~\bibnamefont{Aizenman}} \bibnamefont{and}
  \bibinfo{author}{\bibfnamefont{J.}~\bibnamefont{Wehr}},
  \bibinfo{journal}{Phys.\ Rev.\ Lett} \textbf{\bibinfo{volume}{62}},
  \bibinfo{pages}{2503} (\bibinfo{year}{1989}).

\bibitem[{\citenamefont{Uzelac et~al.}(1995)\citenamefont{Uzelac, Hasmy, and
  Jullien}}]{3}
\bibinfo{author}{\bibfnamefont{K.}~\bibnamefont{Uzelac}},
  \bibinfo{author}{\bibfnamefont{A.}~\bibnamefont{Hasmy}}, \bibnamefont{and}
  \bibinfo{author}{\bibfnamefont{R.}~\bibnamefont{Jullien}},
  \bibinfo{journal}{Phys.\ Rev.\ Lett} \textbf{\bibinfo{volume}{74}},
  \bibinfo{pages}{422} (\bibinfo{year}{1995}).

\bibitem[{\citenamefont{Ballesteros et~al.}(2000)\citenamefont{Ballesteros,
  Fernandez, Martin-Mayor, Sudupe, Parisi, and Ruiz-Lorenzo}}]{4}
\bibinfo{author}{\bibfnamefont{H.~G.} \bibnamefont{Ballesteros}},
  \bibinfo{author}{\bibfnamefont{L.~A.} \bibnamefont{Fernandez}},
  \bibinfo{author}{\bibfnamefont{V.}~\bibnamefont{Martin-Mayor}},
  \bibinfo{author}{\bibfnamefont{A.~M.} \bibnamefont{Sudupe}},
  \bibinfo{author}{\bibfnamefont{G.}~\bibnamefont{Parisi}}, \bibnamefont{and}
  \bibinfo{author}{\bibfnamefont{J.~J.} \bibnamefont{Ruiz-Lorenzo}},
  \bibinfo{journal}{Phys.\ Rev.\ B} \textbf{\bibinfo{volume}{61}},
  \bibinfo{pages}{3215} (\bibinfo{year}{2000}).

\bibitem[{\citenamefont{Chatelain et~al.}(2001)\citenamefont{Chatelain, Berche,
  Janke, and Berche}}]{5}
\bibinfo{author}{\bibfnamefont{C.}~\bibnamefont{Chatelain}},
  \bibinfo{author}{\bibfnamefont{B.}~\bibnamefont{Berche}},
  \bibinfo{author}{\bibfnamefont{W.}~\bibnamefont{Janke}}, \bibnamefont{and}
  \bibinfo{author}{\bibfnamefont{P.-E.} \bibnamefont{Berche}},
  \bibinfo{journal}{Phys.\ Rev.\ E} \textbf{\bibinfo{volume}{64}},
  \bibinfo{pages}{036120} (\bibinfo{year}{2001}).

\bibitem[{\citenamefont{Janke et~al.}()\citenamefont{Janke, Berche, Chatelain,
  and Berche}}]{6}
\bibinfo{author}{\bibfnamefont{W.}~\bibnamefont{Janke}},
  \bibinfo{author}{\bibfnamefont{P.-E.} \bibnamefont{Berche}},
  \bibinfo{author}{\bibfnamefont{C.}~\bibnamefont{Chatelain}},
  \bibnamefont{and} \bibinfo{author}{\bibfnamefont{B.}~\bibnamefont{Berche}},
  \eprint{cond-mat/0304642}.

\bibitem[{\citenamefont{Imry and Ma}(1975)}]{7}
\bibinfo{author}{\bibfnamefont{Y.}~\bibnamefont{Imry}} \bibnamefont{and}
  \bibinfo{author}{\bibfnamefont{S.~K.} \bibnamefont{Ma}},
  \bibinfo{journal}{Phys.\ Rev.\ Lett.} \textbf{\bibinfo{volume}{35}},
  \bibinfo{pages}{1399} (\bibinfo{year}{1975}).

\bibitem[{\citenamefont{Imbrie}(1984)}]{8}
\bibinfo{author}{\bibfnamefont{J.}~\bibnamefont{Imbrie}},
  \bibinfo{journal}{Phys.\ Rev.\ Lett.} \textbf{\bibinfo{volume}{53}},
  \bibinfo{pages}{1747} (\bibinfo{year}{1984}).

\bibitem[{\citenamefont{Cardy and Jacobsen}(1997)}]{9}
\bibinfo{author}{\bibfnamefont{J.}~\bibnamefont{Cardy}} \bibnamefont{and}
  \bibinfo{author}{\bibfnamefont{J.}~\bibnamefont{Jacobsen}},
  \bibinfo{journal}{Phys.\ Rev.\ Lett.} \textbf{\bibinfo{volume}{79}},
  \bibinfo{pages}{4063} (\bibinfo{year}{1997}).

\bibitem[{\citenamefont{Cardy}()}]{10}
\bibinfo{author}{\bibfnamefont{J.}~\bibnamefont{Cardy}},
  \eprint{cond-mat/9806355}.

\bibitem[{\citenamefont{Morgenstern et~al.}(1981)\citenamefont{Morgenstern,
  Binder, and Hornreich}}]{11}
\bibinfo{author}{\bibfnamefont{I.}~\bibnamefont{Morgenstern}},
  \bibinfo{author}{\bibfnamefont{K.}~\bibnamefont{Binder}}, \bibnamefont{and}
  \bibinfo{author}{\bibfnamefont{R.~M.} \bibnamefont{Hornreich}},
  \bibinfo{journal}{Phys.\ Rev.\ B} \textbf{\bibinfo{volume}{23}},
  \bibinfo{pages}{287} (\bibinfo{year}{1981}).

\bibitem[{\citenamefont{Esser et~al.}()\citenamefont{Esser, Novak, and
  Usadel}}]{12}
\bibinfo{author}{\bibfnamefont{J.}~\bibnamefont{Esser}},
  \bibinfo{author}{\bibfnamefont{U.}~\bibnamefont{Novak}}, \bibnamefont{and}
  \bibinfo{author}{\bibfnamefont{K.~D.} \bibnamefont{Usadel}},
  \eprint{cond-mat/9612022}.

\bibitem[{\citenamefont{Seppala and Alava}(2001)}]{13}
\bibinfo{author}{\bibfnamefont{E.~T.} \bibnamefont{Seppala}} \bibnamefont{and}
  \bibinfo{author}{\bibfnamefont{M.~J.} \bibnamefont{Alava}},
  \bibinfo{journal}{Phys.\ Rev.\ E} \textbf{\bibinfo{volume}{63}},
  \bibinfo{pages}{066109} (\bibinfo{year}{2001}).

\bibitem[{\citenamefont{Seppala et~al.}(2002)\citenamefont{Seppala, Pulkkinen,
  and Alava}}]{14}
\bibinfo{author}{\bibfnamefont{E.~T.} \bibnamefont{Seppala}},
  \bibinfo{author}{\bibfnamefont{A.~M.} \bibnamefont{Pulkkinen}},
  \bibnamefont{and} \bibinfo{author}{\bibfnamefont{M.~J.} \bibnamefont{Alava}},
  \bibinfo{journal}{Phys.\ Rev.\ B} \textbf{\bibinfo{volume}{66}},
  \bibinfo{pages}{144403} (\bibinfo{year}{2002}).

\bibitem[{\citenamefont{Bastea and Duxbury}()}]{15}
\bibinfo{author}{\bibfnamefont{S.}~\bibnamefont{Bastea}} \bibnamefont{and}
  \bibinfo{author}{\bibfnamefont{P.~M.} \bibnamefont{Duxbury}},
  \eprint{cond-mat/9801108}.

\bibitem[{\citenamefont{Berche and Chatelain}()}]{16}
\bibinfo{author}{\bibfnamefont{B.}~\bibnamefont{Berche}} \bibnamefont{and}
  \bibinfo{author}{\bibfnamefont{C.}~\bibnamefont{Chatelain}},
  \eprint{cond-mat/0207421}.

\bibitem[{\citenamefont{Dotsenko et~al.}(1998)\citenamefont{Dotsenko, Dotsenko,
  and Picco}}]{17}
\bibinfo{author}{\bibfnamefont{V.}~\bibnamefont{Dotsenko}},
  \bibinfo{author}{\bibfnamefont{V.}~\bibnamefont{Dotsenko}}, \bibnamefont{and}
  \bibinfo{author}{\bibfnamefont{M.}~\bibnamefont{Picco}},
  \bibinfo{journal}{Nucl.\ Phys.\ B} \textbf{\bibinfo{volume}{250}},
  \bibinfo{pages}{633} (\bibinfo{year}{1998}).

\bibitem[{\citenamefont{Zeng et~al.}(1999)\citenamefont{Zeng, Zalar,
  Iannacchione, and Finotello}}]{18}
\bibinfo{author}{\bibfnamefont{H.}~\bibnamefont{Zeng}},
  \bibinfo{author}{\bibfnamefont{B.}~\bibnamefont{Zalar}},
  \bibinfo{author}{\bibfnamefont{G.~S.} \bibnamefont{Iannacchione}},
  \bibnamefont{and}
  \bibinfo{author}{\bibfnamefont{D.}~\bibnamefont{Finotello}},
  \bibinfo{journal}{Phys.\ Rev.\ E} \textbf{\bibinfo{volume}{60}},
  \bibinfo{pages}{5607} (\bibinfo{year}{1999}).

\end{thebibliography}
\end{document}